\documentclass[12pt]{article}
\usepackage{amssymb}
\usepackage{amsmath}
\usepackage{enumerate}
\newtheorem{Proposition}{Proposition}

\newcommand\btd{\raise 2pt \hbox{$\hat\bigtriangledown$}\hskip 1.5pt}
\newcommand\bt{\raise 2pt \hbox{$\bigtriangledown$}\hskip 1.5pt}
\begin{document}

\title{\large\bf Bilinear Approach to $N=2$ Supersymmetric KdV equations }
\author{ Meng-Xia Zhang$^{\dag}$$^\ddag$, Q. P. Liu$^{\dag}$,   Ya-Li Shen$^{\dag}$$^\sharp$, Ke Wu$^{\ddag}$,
\\[10pt]
\footnotesize
 \small $^\dag$Department of Mathematics, \\
\small China University of Mining and Technology,\\
\small Beijing 100083, P. R. China.\\
\small $^\sharp$Yuncheng University, Shanxi 044000, P. R. China.\\
\footnotesize \small $^\ddag$School of Mathematical Sciences, Capital Normal University,\\
\small Beijing 100037, P. R. China.}
\date{}
\maketitle
\begin{abstract}
The $N=2$ supersymmetric KdV equations are studied within the
framework of Hirota's bilinear method. For two such equations,
namely $N=2, a=4$ and $N=2, a=1$ supersymmetric KdV equations, we
obtain the corresponding bilinear formulations. Using them, we
construct particular solutions for both cases. In particular, a
bilinear B\"{a}cklund transformation is given for the $N=2, a=1$
supersymmetric KdV equation.
\end{abstract}

\newpage

\section{Introduction}

The theory of supersymmetric integrable systems has been an
extensive research field for more than twenty years. As a
consequence, many supersymmetric integrable equations have been
studied and a number of interesting properties has been established.
Among them, the most celebrated supersymmetric system is the
supersymmetric Korteweg-de Vries (KdV) equation \cite{MR,MP1}. It
has been shown that, as its bosonic analogue, the supersymmetric KdV
(SKdV) equation is a bi-Hamiltonian system \cite{OP1}, has Darboux
and B\"{a}cklund transformations \cite{LM,LX}, can be casted into
bilinear from \cite{MY,CAS,CRG}, etc.

In the literature, there exist more than one supersymmetric
extensions for the KdV equation. The most interesting ones are the
N=2 supersymmetric KdV equations.  The system was originally
introduced by Laberge and Mathieu \cite{LCM,LPM}. It reads as
\begin{equation}
\label{SKdVa}\phi_t=-\phi_{xxx}+3(\phi{\cal D}_1{\cal
D}_2\phi)_x+\frac{1}{2} (a-1)({\cal D}_1{\cal
D}_2\phi^2)_x+3a\phi^2\phi_x
\end{equation}
where $\phi=\phi(x,t,\theta_1,\theta_2)$ is a superboson function
depending on temporal variable $t$, spatial variable $x$ and its
fermionic counterparts $\theta_i (i=1,2)$. ${\cal D}_1$ and ${\cal
D}_2$ are the super derivatives defined by ${\cal
D}_1=\partial_{\theta_1}+\theta_1\partial_x, {\cal
D}_2=\partial_{\theta_2}+\theta_2\partial_x$ and $a$ is a parameter.
In the sequel, we will refer to (\ref{SKdVa}) as the SKdV$_a$
equation. This one-parameter family of equations (\ref{SKdVa}) is
integrable only for certain values of the parameter $a$. Indeed,
Laberge and Mathieu in \cite{LCM} have shown that for both $a=-2$
and $a=4$, there exist Lax operators and Hamiltonian structures and
infinite conservation laws. Then they \cite{LPM} introduced $SKdV_1$
equation as a Hamiltonian equation with the $N=2$ superconformal
algebra as a second Hamiltonian structure. The Lax representations
are given for SKdV$_{-2}$ and SKdV$_1$ equations in \cite{Pop}.
Kupershmidt (see ref. \cite{LPM}) observed that SKdV$_4$ equation is
actually a bi-Hamiltonian system while Oevel and Popowicz \cite{OP1}
constructed the bi-Hamiltonian structures for both  SKdV$_{-2}$ and
SKdV$_4$ equations based on $r$-matrix theory. A Painlev\'{e}
analysis is performed for the SKdV$_a$ equations in \cite{BM}. We
also remark that $N=2$ SKdV systems can be represented in terms of
$N=1$ Lax operators \cite{IK,Liu2}.

The purpose of this paper is to study the SKdV$_a$ equation from the
viewpoint of Hirota's method. Recently, Hirota's method has been
applied to the supersymmetric integrable systems and the equations
considered includes SKdV equation \cite{CRG, LH}, supersymmetric
MKdV equation \cite{LHZ}, supersymmetric classical Boussinesq
equation or supersymmetric two-boson equation \cite{LY}. As in the
classical case, Hirota's method can be adopted not only for
constructing solutions, but also can be used for derivations of
other properties. Therefore, this approach is very effective in the
study of supersymmetric systems.

The paper is organized as follows. In section 2, we will transform
the SKdV$_4$ equation into bilinear form and construct its solitons.
In section 3, we first convert the SKdV$_1$ equation into  bilinear
form, then making use of this bilinear form a B\"{a}cklund
transformation and soliton solutions are constructed for this
system. And a Lax representation can be worked out for the SKdV$_1$
equation. Final section contains a brief discussion.

\section{SKdV$_4$ equation}
In this section, we will consider the SKdV$_4$ and show that it
can be converted into a bilinear form. Our strategy to do so is
first to rewrite this equation in terms of N=1 formalism, then we
embed it into the hierarchy of the supersymmetric two-boson
system. Based on this connection, we provide a proper bilinear
form for the SKdV$_4$. We also construct soliton solutions for
this system.

\subsection{Bilinear form }
From  (\ref{SKdVa}), our SKdV$_4$ equation reads as
\begin{equation}
\phi_t=\left[-\phi_{xx}+3\phi{\cal D}_1{\cal
D}_2\phi+\frac{3}{2}{\cal D}_1{\cal D}_2\phi^2+4\phi^3\right]_x,
\label{SKdV4}
\end{equation}
let
\[
\phi=v+\theta_2\beta
\]
where $v=v(t,x,\theta_1)$ is a bosonic (even) function while
$\beta=\beta(t,x,\theta_1)$ is a fermionic (odd) one. Then the
SKdV$_4$ equation (\ref{SKdV4}) in components takes the following
form
\begin{subequations}
\begin{eqnarray}
\label{3a} v_t&=&\left[-v_{xx}+6v{\cal D}\beta+3({\cal D}v)\beta+4v^3\right]_x, \\
\label{3b}\beta_t&=&\left[-\beta_{xx}-6v{\cal D}v_x-3v_x{\cal
D}v+3\beta{\cal D}\beta+12v^2\beta\right]_x,
\end{eqnarray}
\end{subequations}
where and hereafter we use ${\cal D}={\cal D}_1$ for simplicity.
Suppose that
\[
 v=\frac{1}{2}iu,\quad \beta=-\alpha+\frac{1}{2}{\cal
D}u
\]
where $i=\sqrt{-1}$, then the system (\ref{3a}-\ref{3b}) is
transformed into
\begin{subequations}
\begin{eqnarray}
u_t&=&\left[-u_{xx}+3\alpha{\cal D}u-6u{\cal
D}\alpha-u^3+3uu_x\right]_x,\label{4a}\\
\alpha_t&=&\left[-\alpha_{xx}-3\alpha{\cal
D}\alpha-3u^2\alpha-3u\alpha_x\right]_x.\label{4b}
\end{eqnarray}
\end{subequations}
It has been shown that above system  and the supersymmetric
two-boson (sTB) equation share the  same hierarchy \cite{Liu2}. The
system (\ref{4a}-\ref{4b}) in fact is the third flow while the sTB
is the second one. So we recall the sTB equation \cite{BD}
\begin{subequations}
\begin{eqnarray}
u_{t_2}&=&(-u_x+u^2+2{\cal D}\alpha)_x,\label{5a}\\
\alpha_{t_2}&=&(\alpha_x+2u\alpha)_x.\label{5b}
\end{eqnarray}
\end{subequations}
As shown in \cite{LY}, through the following dependent variables
transformations
\begin{equation}
u=-\left(\ln\frac{f}{g}\right)_x=-\varphi_x, \qquad \alpha=({\cal
D}\ln g)_x={\cal D}\rho_x,\label{tran}
\end{equation}
the sTB equation is brought into the  bilinear form
\begin{subequations}
\begin{eqnarray}
(D_{t_2}+D_x^2)f\cdot g&=&0,\label{6a}\\
S(D_{t_2}+D_x^2)f\cdot g&=&0,\label{6b}
\end{eqnarray}
\end{subequations}
where the super Hirota derivative is defined as:
\[
SD_t^mD_x^n f\cdot g=({\cal D}_{\theta_1}-{\cal
D}_{\theta_2})\left({\partial\over\partial
t_1}-{\partial\over\partial
t_2}\right)^m\left({\partial\over\partial
x_1}-{\partial\over\partial x_2}\right)^n
f(x_1,t_1,\theta_1)g(x_2,t_2,\theta_2)\left|_{\substack {x_1=x_2=x\\
t_1=t_2=t\\ \theta_1=\theta_2=\theta}}.\right.
\]

Now substituting the transformations (\ref{tran}) into the eqs.
(\ref{5a}) and (\ref{5b}), we obtain
\begin{subequations}
\begin{eqnarray}
\varphi_{t_2}&=&-\varphi_{xx}-\varphi_x^2-2\rho_{xx},\label{8a}\\
\rho_{t_2}&=&\rho_{xx}-2{\cal D}^{-1}(\varphi_x{\cal
D}\rho_x).\label{8b}
\end{eqnarray}
\end{subequations}
Observing that the eqs. (\ref{5a}-\ref{5b}) or (\ref{8a}-\ref{8b})
and the eqs. (\ref{4a}-\ref{4b}) are the second and third flows
respectively, we now convert the latter into bilinear form.

With the help of the transformations (\ref{tran}), the system
(\ref{4a})-(\ref{4b}) is rewritten as
\begin{eqnarray*}
[\varphi_{t}+\varphi_{xxx}+3({\cal D}\varphi_x)({\cal
D}\rho_x)+6\varphi_x\rho_{xx}+\varphi_x^3+3\varphi_x\varphi_{xx}]_x&=&0,\\
({\cal D}\rho_{t}+{\cal D}\rho_{xxx}+3\varphi_x^2{\cal
D}\rho_x-3\varphi_x{\cal D}\rho_{xx}+3\rho_{xx}{\cal
D}\rho_x)_x&=&0,
\end{eqnarray*}
integrating the above equations once and taking zero as the
integration constants, we obtain
\begin{subequations}
\begin{eqnarray}
\varphi_{t}+\varphi_{xxx}+3({\cal D}\varphi_x)({\cal
D}\rho_x)+6\varphi_x\rho_{xx}+\varphi_x^3+3\varphi_x\varphi_{xx}&=&0,\label{9a}\\
{\cal D}\rho_{t}+{\cal D}\rho_{xxx}+3\varphi_x^2{\cal
D}\rho_x-3\varphi_x{\cal D}\rho_{xx}+3\rho_{xx}{\cal
D}\rho_x&=&0.\label{9b}
\end{eqnarray}
\end{subequations}
For (\ref{9a}), we have
\begin{eqnarray}
0&=&\varphi_{t}+\frac{1}{4}(\varphi_x^3+3\varphi_x\varphi_{xx}+6\varphi_x\rho_{xx}+\varphi_{xxx})\nonumber\\
&&+\frac{3}{4}[\varphi_x^3+3\varphi_x\varphi_{xx}+6\varphi_x\rho_{xx}+\varphi_{xxx}+4({\cal
D}\varphi_x)({\cal
D}\rho_x)]\nonumber\\
&=&\varphi_{t}+\frac{1}{4}(\varphi_x^3+3\varphi_x\varphi_{xx}+6\varphi_x\rho_{xx}+\varphi_{xxx})\nonumber\\
&&-\frac{3}{4}\left\{\varphi_x(-\varphi_{xx}-\varphi_x^2-2\rho_{xx})+(-\varphi_{xx}-\varphi_x^2-2\rho_{xx})_x+2[\rho_{xx}-2{\cal
D}^{-1}(\varphi_x{\cal
D}\rho_x)]_x\right\}\nonumber\\
&\overset{(\ref{8a}),(\ref{8b})}{=}&\varphi_{t}+\frac{1}{4}(\varphi_x^3+3\varphi_x\varphi_{xx}+6\varphi_x\rho_{xx}+\varphi_{xxx})-\frac{3}{4}(\varphi_x\varphi_{t_2}+\varphi_{xt_2}+2\rho_{xt_2})\nonumber\\
&=&\frac{1}{fg}(D_{t}-\frac{3}{4}D_xD_{t_2}+\frac{1}{4}D_x^3)f\cdot
g\label{10}
\end{eqnarray}
and for (\ref{9b}), we have
\begin{eqnarray}
0&=&\frac{1}{2}\left\{-[\varphi_x^3+3\varphi_x\varphi_{xx}+6\varphi_x\rho_{xx}+\varphi_{xxx}+3({\cal
D}\varphi_x)({\cal D}\rho_x)]{\cal D}\varphi\right.\nonumber\\
&&-{\cal
D}[\varphi_x^3+3\varphi_x\varphi_{xx}+6\varphi_x\rho_{xx}+\varphi_{xxx}+3({\cal
D}\varphi_x)({\cal D}\rho_x)]+2{\cal D}\rho_{t}\nonumber\\
&&+\frac{3}{4}[\varphi_x^3+3\varphi_x\varphi_{xx}+6\varphi_x\rho_{xx}+\varphi_{xxx}+4({\cal
D}\varphi_x)({\cal D}\rho_x)]{\cal
D}\varphi\nonumber\\
&&+\frac{3}{4}(\varphi_{xx}+\varphi_x^2+2\rho_{xx}){\cal
D}\varphi_x+\frac{3}{4}\varphi_x{\cal
D}(\varphi_{xx}+\varphi_x^2+2\rho_{xx})\nonumber\\
&&-\frac{3}{2}\varphi_x{\cal D}[\rho_{xx}-2{\cal
D}^{-1}(\varphi_x{\cal
D}\rho_x)]+\frac{3}{2}(\varphi_{xx}+\varphi_x^2+2\rho_{xx}){\cal
D}\rho_x\nonumber\\
&&+\frac{3}{4}{\cal
D}(\varphi_{xx}+\varphi_x^2+2\rho_{xx})_x+\frac{1}{4}[(\varphi_x^3+3\varphi_x\varphi_{xx}+6\varphi_x\rho_{xx}+\varphi_{xxx}){\cal
D}\varphi\nonumber\\
&&+3\varphi_x{\cal D}\varphi_{xx}+{\cal D}\varphi_{xxx}+2{\cal
D}\rho_{xxx}+3(\varphi_{xx}+\varphi_x^2+2\rho_{xx}){\cal
D}\varphi_x\nonumber\\
&&\left.+6(\varphi_{xx}+\varphi_x^2+2\rho_{xx}){\cal
D}\rho_x]\right\}\nonumber\\
&\overset{(\ref{8a}-\ref{9a})}{=}&\frac{1}{2}\left\{\varphi_{t}{\cal
D}\varphi+{\cal D}\varphi_{t}+2{\cal
D}\rho_{t}-\frac{3}{4}[(\varphi_x\varphi_{t_2}+\varphi_{xt_2}+2\rho_{xt_2}){\cal
D}\varphi\right.\nonumber\\
&&\left.+\varphi_{t_2}{\cal D}\varphi_x+\varphi_x{\cal
D}\varphi_{t_2}+2\varphi_x{\cal D}\rho_{t_2}+2\varphi_{t_2}{\cal
D}\rho_x+{\cal D}\varphi_{xt_2}]+\frac{1}{4}\frac{SD_x^3f\cdot g}{fg}\right\}\nonumber\\
&=&\frac{1}{2fg}S\left(D_{t}-\frac{3}{4}D_xD_{t_2}+\frac{1}{4}D_x^3\right)f\cdot
g\label{11}
\end{eqnarray}
therefore, our  SKdV$_4$ equation assumes the following bilinear
form
\begin{subequations}
\begin{eqnarray}
(D_{t}-\frac{3}{4}D_xD_{t_2}+\frac{1}{4}D_x^3)f\cdot g&=&0,\label{bi1}\\
S(D_{t}-\frac{3}{4}D_xD_{t_2}+\frac{1}{4}D_x^3)f\cdot
g&=&0.\label{bi2}
\end{eqnarray}
\end{subequations}

\subsection{Solutions}

For a given system, Hirota's bilinear form is ideal for constructing
particular solutions. Next we shall show that  a class of solutions
can be calculated for the $SKdV_4$ equation. We take
\begin{equation*}
f=\varepsilon f_1, \qquad g=1+\varepsilon g_1+\varepsilon^2
g_2+\varepsilon^3 g_3+\cdots,
\end{equation*}
substituting the above expressions into the eqs. (\ref{bi1}) and
(\ref{bi2}) and collecting the alike power terms, we have
\begin{equation}
  \begin{array}{ccr}
    \varepsilon^1: & & (D_{t}-\frac{3}{4}D_xD_{t_2}+\frac{1}{4}D_x^3)(f_1\cdot
1)=0, \\[10pt]
     &  & S(D_{t}-\frac{3}{4}D_xD_{t_2}+\frac{1}{4}D_x^3)(f_1\cdot
1)=0, \\
  \end{array}\label{13a}
\end{equation}
and for $i\geq1$,
\begin{equation}
  \begin{array}{ccc}
    \varepsilon^{i+1}: &   & D_{t}-\frac{3}{4}D_xD_{t_2}+\frac{1}{4}D_x^3)(f_1\cdot
g_i)=0, \\[10pt]
     &  &  S(D_{t}-\frac{3}{4}D_xD_{t_2}+\frac{1}{4}D_x^3)(f_1\cdot
g_i)=0.\\
  \end{array}\label{14}
\end{equation}

From (\ref{13a}), we get
\begin{equation}
f_{1t}-\frac{3}{4}f_{1xt_2}+\frac{1}{4}f_{1xxx}=0,\quad \quad {\cal
D}(f_{1t}-\frac{3}{4}f_{1xt_2}+\frac{1}{4}f_{1xxx})=0.\label{15}
\end{equation}
From eqs. (\ref{6a}) and (\ref{6b}), we may take \cite{LY},
\begin{equation}
f_{1t_2}=-f_{1xx},\quad\quad g_{i,t_2}=g_{i,xx}-2kg_{i,x}.\label{16}
\end{equation}
substituting the above expression into (\ref{15}), we get
$f_{1t}=-f_{1xxx}$. Therefor, $f_1$ assumes the following form
\begin{equation}
f_1=e^{kx-k^2t_2-k^3t+\theta\xi}\label{17}
\end{equation}
where $k$ is an usual constant and $\xi$ is a Grassmann odd
constant. Then, substituting eqs. (\ref{16}) and (\ref{17}) into
(\ref{14}), we obtain
\begin{eqnarray*}
g_{i,t}+3k^2g_{i,x}-3kg_{i,xx}+g_{i,xxx}&=&0,\\
({\cal D}-\xi-\theta
k)(g_{i,t}+3k^2g_{i,x}-3kg_{i,xx}+g_{i,xxx})&=&0.
\end{eqnarray*}
Accordingly we can choose
\[g_i=e^{k_ix+k_i(k_i-2k)t_2+k_i(3kk_i-k_i^2-3k^2)t+\theta\xi_i}\]
so, we have
\begin{equation}
f=e^{kx-k^2t_2-k^3t+\theta\xi},\quad g=1+\sum_{i=1}^N
e^{k_ix+k_i(k_i-2k)t_2+k_i(3kk_i-k_i^2-3k^2)t+\theta\xi_i}.
\end{equation}
These solutions have the remarkable property that they allow
fusion and fission to take place \cite{MY,CAS}.

\section{SKdV$_1$ equation}
In above section, we succeeded to construct the bilinear form and a
class of solutions for the SKdV$_4$ equation. Now we turn to the
SKdV$_1$ equation and study it from the viewpoint of Hirota's
bilinear method. We will show that this system enjoys a simple
bilinear form and a remarkable B\"{a}cklund transformation.

\subsection{Bilinear form}
As in the case of the SKdV$_4$ equation, we will work in the context
of $N=1$ formalism. Therefore, let
\[
\phi=v+\theta_2\beta
\]
then from the system (\ref{SKdVa}), our SKdV$_1$ equation in
component reads as
\begin{subequations}
\begin{eqnarray}
\label{21a}v_t&=&[-v_{xx}+3v{\cal D}\beta+v^3]_x, \\
\label{21b}\beta_t&=&[-\beta_{xx}-3v{\cal D}v_x+3\beta{\cal
D}\beta+3v^2\beta]_x.
\end{eqnarray}
\end{subequations}
Now we introduce the following dependent variable transformations
\begin{equation}
v=i\left(\ln\frac{f}{g}\right)_x=i\varphi_x, \qquad \beta=-({\cal
D}\ln fg)_x=-{\cal D}\rho_x,\label{tran2}
\end{equation}
substituting the above expressions into the eqs.
(\ref{21a}-\ref{21b}), we obtain
\begin{subequations}
\begin{eqnarray}
\label{23a}\varphi_t+\varphi_{xxx}+3\varphi_x\rho_{xx}+\varphi_x^3&=&0,\\
\label{23b}{\cal D}\rho_t+{\cal D}\rho_{xxx}+3\varphi_x{\cal
D}\varphi_{xx}+3\rho_{xx}{\cal D}\rho_x+3\varphi_x^2{\cal
D}\rho_x&=&0,
\end{eqnarray}
\end{subequations}
for (\ref{23a}), we have
\begin{equation}
\varphi_t+\varphi_x^3+\varphi_{xxx}+3\varphi_x\rho_{xx}=\frac{1}{fg}(D_t+D_x^3)f\cdot
g=0,\label{24}
\end{equation}
and for (\ref{23b}), we have
\begin{eqnarray}
0&=&{\cal D}\rho_t+{\cal D}\rho_{xxx}+3\varphi_x{\cal
D}\varphi_{xx}+3\rho_{xx}{\cal D}\rho_x+3\varphi_x^2{\cal
D}\rho_x\nonumber\\
&\overset{(\ref{23a})}{=}&{\cal D}\rho_t+{\cal
D}\rho_{xxx}+3\varphi_x{\cal D}\varphi_{xx}+3\rho_{xx}{\cal
D}\rho_x+3\varphi_x^2{\cal
D}\rho_x+(\varphi_t+\varphi_{xxx}+3\varphi_x\rho_{xx}+\varphi_x^3){\cal
D}\varphi\nonumber\\
&=&\frac{1}{fg}S(D_t+D_x^3)f\cdot g.\label{25}
\end{eqnarray}
From the above eqs. (\ref{24}-\ref{25}), we obtain the bilinear form
for the SKdV$_1$ equation
\begin{subequations}
\begin{eqnarray}
(D_t+D_x^3)f\cdot g&=&0,\label{1bi1}\\
S(D_t+D_x^3)f\cdot g&=&0.\label{1bi2}
\end{eqnarray}
\end{subequations}

\noindent
{\bf Remark}: This bilinerization is particularly simple
and can be considered as a direct generalization of the bilinear
form of the supersymmetric two-boson system (\ref{6a}-\ref{6b}).
However, it is interesting that these two systems do not belong to
the same integrable hierarchy.

\subsection{B\"{a}cklund transformation}
Integrable systems often possess B\"{a}cklund transformations, which
may be used to construct solutions. Also,  B\"{a}cklund
transformation is  considered as  a characteristic of integrability
for a given system. In this section, we will derive a bilinear BT
for the SKdV$_1$ system. We follow the paper \cite{LHZ} and our
results are summarized in the following

\begin{Proposition} Suppose that $(f,g)$ is a solution of eqs.
(\ref{1bi1}) and (\ref{1bi2}), then $(f',g')$ satisfying the
following relations,
\begin{subequations}
\begin{eqnarray}
\label{Ba1}D_x g\cdot f'-D_x f\cdot g'=\mu gf'-\mu fg',\\
\label{Ba2}SD_x g\cdot f'+SD_x f\cdot g'=\mu S g\cdot f'+\mu S
f\cdot g',\\
\label{Ba3}(D_t+D_x^3-3\mu D_x^2+3\mu^2D_x)g\cdot g'=0,\\
\label{Ba4}(D_t+D_x^3-3\mu D_x^2+3\mu^2D_x)f\cdot f'=0,
\end{eqnarray}
\end{subequations}
is another solution of (\ref{1bi1}) and (\ref{1bi2}), where $\mu$ is
an ordinary constant.
\end{Proposition}

\noindent{\bf Proof.} We consider the following
\begin{eqnarray*}
\mathbb{P}_1&\equiv&[(D_t+D_x^3)f\cdot g]f'g'-fg[(D_t+D_x^3)f'\cdot g'],\\
\mathbb{P}_2&\equiv&[S(D_t+D_x^3)f\cdot
g]f'g'-fg[S(D_t+D_x^3)f'\cdot g'].
\end{eqnarray*}
We will show that above eqs. (\ref{Ba1})-(\ref{Ba4}) imply
$\mathbb{P}_1=0$ and $\mathbb{P}_2=0$. The case of $\mathbb{P}_1$
can be verified as in \cite{LHZ}, so we will concentrate on
$\mathbb{P}_2$ next. We will use various bilinear identities which
are presented in Appendix A.
\begin{eqnarray*}
\mathbb{P}_2&\overset{(\ref{A.1},\ref{A.2})}{=}&S[(D_t g\cdot
g')\cdot ff'-gg'\cdot(D_t f\cdot f')]+(S f\cdot g)(D_t f'\cdot
g')-(D_t
f\cdot g)(S f'\cdot g')\\
&&-3D_x[(SD_x f\cdot g')\cdot (D_x g\cdot f')+(SD_x g\cdot
f')\cdot(D_x f\cdot g')]+(S f\cdot g)(D_x^3 f'\cdot g')\\
&&-(D_x^3 f\cdot g)(S f'\cdot g')+S[(D_x^3 f\cdot f')\cdot
gg'-ff'\cdot(D_x^3 g\cdot g')]\\
&=&-3D_x[(SD_x f\cdot g')\cdot(D_x g\cdot f')+(SD_x g\cdot
f')\cdot(D_x f\cdot g')]\\
&&+S\{[(D_t+D_x^3)g\cdot g']\cdot ff'+[(D_t+D_x^3)f\cdot f']\cdot
gg'\}\\
&&+(S f\cdot g)[(D_t+D_x^3)f'\cdot g']-(S f'\cdot g')[(D_t+D_x^3) f\cdot g]\\
&\overset{(\ref{Ba3},\ref{Ba4})}{=}&-3D_x[(SD_x f\cdot g')\cdot(D_x
g\cdot f')+(SD_x g\cdot f')\cdot(D_x f\cdot g')]\\
&&+3\mu S[(D_x^2 g\cdot g')\cdot ff'-gg'\cdot(D_x^2 f\cdot
f')]-3\mu^2 S[(D_x g\cdot g')\cdot
ff'-gg'\cdot (D_x f\cdot f')]\\
&&+(S f\cdot g)[(D_t+D_x^3)f'\cdot g']-(S f'\cdot g')[(D_t+D_x^3) f\cdot g]\\
&\overset{(\ref{A.3},\ref{A.4})}{=}&-3D_x[(SD_x f\cdot g')\cdot(D_x
g\cdot f')+(SD_x g\cdot f')\cdot(D_x f\cdot g')]\\
&&+3\mu D_x[(SD_x g\cdot f')\cdot fg'+(D_x g\cdot f')\cdot(S g'\cdot
f)-(S g\cdot f')\cdot(D_x g'\cdot f)\\
&&-gf'\cdot(SD_x g'\cdot f)]-3\mu^2 D_x[(S g\cdot f')\cdot fg'+gf'\cdot (S g'\cdot f)]\\
&&+(S f\cdot g)[(D_t+D_x^3)f'\cdot g']-(S f'\cdot g')[(D_t+D_x^3) f\cdot g]\\
&\overset{(\ref{Ba1})}{=}&-3D_x[(SD_x f\cdot g')\cdot(D_x f\cdot
g'+\mu gf'-\mu fg')+(SD_x g\cdot f')(D_x g\cdot f'-\mu gf'+\mu
fg')]\\
&&+3\mu D_x[(SD_x g\cdot f')\cdot fg'+(D_x f\cdot g'+\mu gf'-\mu
fg')\cdot (S g'\cdot f)\\
&&+(S g\cdot f')\cdot(D_x g\cdot f'-\mu gf'+\mu fg')-gf'\cdot(SD_x
g'\cdot f)]\\
&&-3\mu^2 D_x[(S g\cdot f')\cdot fg'+gf'\cdot (S g'\cdot f)]\\
&&+(S f\cdot g)[(D_t+D_x^3)f'\cdot g']-(S f'\cdot g')[(D_t+D_x^3) f\cdot g]\\
&=&3D_x[(D_x f\cdot g'-\mu f g') \cdot(SD_x-\mu S)f\cdot g']\\
&&+3D_x[(D_x g\cdot f'-\mu g f') \cdot(SD_x-\mu S)g\cdot f']\\
&&+(S f\cdot g)[(D_t+D_x^3)f'\cdot g']-(S f'\cdot g')[(D_t+D_x^3) f\cdot g]\\
&\overset{(\ref{Ba1})}{=}&3D_x[(D_x g\cdot f'-\mu gf')\cdot(SD_x
f\cdot g'-\mu S f\cdot g'+SD_x g\cdot f'-\mu S g\cdot f')]\\
&&+(S f\cdot g)[(D_t+D_x^3)f'\cdot g']-(S f'\cdot g')[(D_t+D_x^3) f\cdot g]\\
&\overset{(\ref{Ba2})}{=}&(S f\cdot g)[(D_t+D_x^3)f'\cdot g']-(S
f'\cdot g')[(D_t+D_x^3) f\cdot g].
\end{eqnarray*}
Since $(f,g)$ is a solution, it satisfies $(D_t+D_x^3)f\cdot g=0$.
Also, taking account of $\mathbb{P}_1=0$, it yields that
$(D_t+D_x^3)f'\cdot g'=0$. Therefore, we finally have
$\mathbb{P}_2=0$ and the proof is completed.

Next we will demonstrate that a spectral problem can be derived from
the above BT. For this purpose, we assume
 \[m=f'/f,\quad n=g'/g,\]
 then by simple manipulation, from eqs. (\ref{Ba1}-\ref{Ba4}) we have
\begin{subequations}
\begin{eqnarray}
\label{26a}m_x+m \varphi_x+n \varphi_x-n_x+\mu m-\mu n=0,\\
m {\cal D}\varphi_x+2\varphi_x {\cal D}n+n {\cal D}\varphi_x-2 {\cal
D}n
_x-2\mu {\cal D}n-m {\cal D}\rho_x-n {\cal D}\rho_x=0,\\
n_t-3 n_x \varphi_{xx}+3 n_x \rho_{xx}+n_{xxx}-3\mu n
\varphi_{xx}+3\mu n
\rho_{xx}+3\mu n_{xx}+3\mu^2 n_x=0,\\
\label{26d}m_t+3 m_x \varphi_{xx}+3 m_x \rho_{xx}+m_{xxx}+3\mu m
\varphi_{xx}+3\mu m \rho_{xx}+3\mu m_{xx}+3\mu^2 m_x=0.
\end{eqnarray}
\end{subequations}
\begin{equation}
{\cal D}(m+n)_x+\mu{\cal D}(m+n)+(m+n){\cal D}\rho_x+\varphi_x{\cal
D}(m-n)+\mu (m-n) {\cal D}\varphi+(m-n)_x {\cal D}\varphi+(m+n)
\varphi_x {\cal D}\varphi=0.
\end{equation}
 To obtain a more compact form, we introduce
\[U=m-n,\quad V=m+n,\]
in these variables, and the eqs. (\ref{26a}-\ref{26d}) can be
rewritten simply as
\begin{subequations}
\begin{eqnarray}
\label{Lax1}U_x+\varphi_x V+\mu U=0,\\
\varphi_x {\cal D}U+{\cal D}V_x+\mu {\cal D}V+V {\cal D}\rho_x=0,\\
V_t+3\mu \rho_{xx} V+3\mu \varphi_{xx} U+V_{xxx}+3 \varphi_{xx}
U_x+3\rho_{xx} V_x
+3\mu V_{xx}+3\mu^2 V_x=0,\\
U_t+3\mu \rho_{xx} U+3\mu \varphi_{xx} V+U_{xxx}+3 \varphi_{xx}
\label{Lax4}V_x+3\rho_{xx} U_x +3\mu U_{xx}+3\mu^2 U_x=0.
\end{eqnarray}
\end{subequations}
Therefore we have the following
\begin{Proposition} The compatibility condition of
(\ref{Lax1})-(\ref{Lax4}) are the $SKdV_1$ equtions (\ref{21a}) and
(\ref{21b}).
\end{Proposition}
\noindent{\bf Proof:} Direct calculations.

\subsection{Solutions}

Since our SKdV$_1$ system (\ref{21a}-\ref{21b}) has
(\ref{1bi1}-\ref{1bi2}) as its Hirota's bilinear form, we may adopt
the standard perturbation method to find its possible soliton-like
solutions. By tedious but straightforward calculation, we find
one-soliton, two-soliton and three-soliton solutions, which are
listed in the following

\noindent {\underline{One-soliton}}:
\begin{eqnarray*}
f&=&1+e^{\eta+\theta\xi},\\
g&=&1-e^{\eta+\theta\xi}.
\end{eqnarray*}
where $\eta=kx-k^3t+c_0$.

\noindent {\underline{Two-soliton}}:
\begin{eqnarray*}
f&=&1+e^{\eta_1+\theta\xi_1}+e^{\eta_2+\theta\xi_2}+A_{12}e^{\eta_1+\eta_2+\theta(\xi_1+\xi_2)},\\
g&=&1-e^{\eta_1+\theta\xi_1}-e^{\eta_2+\theta\xi_2}+A_{12}e^{\eta_1+\eta_2+\theta(\xi_1+\xi_2)}.
\end{eqnarray*}

\noindent {\underline{Three-soliton}}:
\begin{eqnarray*}
f&=&1+e^{\eta_1+\theta\xi_1}+e^{\eta_2+\theta\xi_2}+e^{\eta_3+\theta\xi_3}\\
&&+A_{12}e^{\eta_1+\eta_2+\theta(\xi_1+\xi_2)}+A_{13}e^{\eta_1+\eta_3+\theta(\xi_1+\xi_3)}+A_{23}e^{\eta_2+\eta_3+\theta(\xi_2+\xi_3)}\\
&&+(m_{13}m_{23}A_{12}+m_{12}m_{32}A_{13}+m_{12}m_{13}A_{23})e^{\eta_1+\eta_2+\eta_3+\theta(\xi_1+\xi_2+\xi_3)},\\
g&=&1-e^{\eta_1+\theta\xi_1}-e^{\eta_2+\theta\xi_2}-e^{\eta_3+\theta\xi_3}\\
&&+A_{12}e^{\eta_1+\eta_2+\theta(\xi_1+\xi_2)}+A_{13}e^{\eta_1+\eta_3+\theta(\xi_1+\xi_3)}+A_{23}e^{\eta_2+\eta_3+\theta(\xi_2+\xi_3)}\\
&&-(m_{13}m_{23}A_{12}+m_{12}m_{32}A_{13}+m_{12}m_{13}A_{23})e^{\eta_1+\eta_2+\eta_3+\theta(\xi_1+\xi_2+\xi_3)}.
\end{eqnarray*}
where $ \eta_i=k_ix-k_i^3t+c_i, m_{ij}={{k_i-k_j}\over {k_i+k_j}}$
and
\[
A_{ij}=\left({{k_i-k_j}\over{k_i+k_j}}\right)\left({{k_j-k_i+2\xi_i\xi_j}\over{k_i+k_j}}
+2\theta{{k_i\xi_j-k_j\xi_i}\over{k_i+k_j}}\right).
\]

\section{Discussions}
In this paper, we study the N=2 SKdV equations within the framework
of the Hirota bilinear method. For two of the three integrable
cases, namely SKdV$_4$ and SKdV$_1$ equations, we succeed in
obtaining their bilinear forms. We also construct the solutions for
both equations and find a simple B\"{a}cklund transformation for the
SKdV$_1$ equation.

Our results are presented in the N=1 form, so it is interesting to
find if it is possible to study these N=2 equations within the N=2
form. Also, there is another N=2 SKdV equation--SKdV$_{-2}$ and
working out its bilinear form is an open problem.

\bigskip

{\bf Acknowledgements}. We should like to thank Xing-Biao Hu and
Sen-Yue Lou for interesting discussions. This work is supported by
the National Natural Science Foundation of China with grant numbers
10671206 and 10231050.

\renewcommand{\theequation}{A.\arabic{equation}}
\setcounter{equation}{0}
\section*{Appendix: some bilinear identities}

In this appendix, some relevant bilinear identities are listed.
Proofs of these identities are straightforward so are omitted. Here
a, b, c and d are arbitrary even functions of the independent
variables x, t, and $\theta$.
\begin{appendix}
\begin{eqnarray}
(SD_x a\cdot b)cd-ab(SD_x c\cdot d)&=&S[(D_x b\cdot d)\cdot
ac-bd\cdot(D_x a\cdot c)]\nonumber\\
&&+(S a\cdot b)(D_x c\cdot d)-(D_x a\cdot b)(S c\cdot
d),\label{A.1}\\
(SD_x^3 a\cdot b)cd-ab(SD_x^3 c\cdot d)&=&-3D_x[(SD_x a\cdot d)\cdot
(D_x b\cdot c)+(SD_x b\cdot c)\cdot(D_x a\cdot d)]\nonumber\\
&&+S[(D_x^3 a\cdot c)\cdot bd-ac\cdot (D_x^3 b\cdot d)]\nonumber\\
&&+(S a\cdot b)(D_x^3 c\cdot d)-(D_x^3 a\cdot b)(S c\cdot
d),\label{A.2}\\
S[(D_x^2 a\cdot b)\cdot cd-ab\cdot(D_x^2 c\cdot d)]&=&D_x[(SD_x
a\cdot c)\cdot bd+(D_x a\cdot c)\cdot (S b\cdot d)]\nonumber\\
&&+D_x[-(S a\cdot c)\cdot(D_x b\cdot d)-ac\cdot (SD_x b\cdot
d)],\label{A.3}\\
S[(D_x a\cdot b)\cdot cd-ab\cdot (D_x c\cdot d)]&=&D_x[(S a\cdot
d)\cdot bc+ad\cdot (S b\cdot c)].\label{A.4}
\end{eqnarray}
\end{appendix}

\end{document}